\documentclass[conference]{IEEEtran}
\usepackage{cite}
\usepackage{amsmath,amssymb,amsfonts}
\usepackage{algorithmic}
\usepackage{graphicx}
\usepackage{textcomp}
\usepackage{xcolor}
\usepackage{dirtytalk} 

\makeatletter
\def\bstctlcite{\@ifnextchar[{\@bstctlcite}{\@bstctlcite[@auxout]}}
\def\@bstctlcite[#1]#2{\@bsphack
  \@for\@citeb:=#2\do{%
    \edef\@citeb{\expandafter\@firstofone\@citeb}%
    \if@filesw\immediate\write\csname #1\endcsname{\string\citation{\@citeb}}\fi}%
  \@esphack}
\makeatother 

\def\BibTeX{{\rm B\kern-.05em{\sc i\kern-.025em b}\kern-.08em
    T\kern-.1667em\lower.7ex\hbox{E}\kern-.125emX}}
    
\begin{document}
\bstctlcite{IEEEexample:BSTcontrol} 

\title{A Notification Based Nudge for Handling Excessive Smartphone Use\\
}


\author{\IEEEauthorblockN{Partha Sarker\IEEEauthorrefmark{1}}
\IEEEauthorblockA{\textit{Computer
Science and Engineering
} \\
Shahjalal University of
Science 
\\and Technology 
\\Sylhet, Bangladesh \\
parthasarker3@gmail.com}
\and
\IEEEauthorblockN{Dipto Dey\IEEEauthorrefmark{1}}
\IEEEauthorblockA{\textit{Computer
Science and Engineering
} \\
Shahjalal University of
Science
\\and Technology
\\Sylhet, Bangladesh \\
dipdey093@gmail.com}
\and
\IEEEauthorblockN{Marium-E-Jannat Mukta}
\IEEEauthorblockA{\textit{Department} \\
University of British Columbia\\
British Columbia, Canada \\
jannat.16.11@gmail.com}
}

\maketitle
\begingroup\renewcommand\thefootnote{\IEEEauthorrefmark{1}}
\footnotetext{Equal contribution}
\endgroup

\begin{abstract}
Excessive use of smartphones is a worldwide known issue. In this study, we proposed a notification-based intervention approach to reduce smartphone overuse without making the user feel any annoyance or irritation. Most of the work in this field tried to reduce smartphone overuse by making smartphone use more difficult for the user. In our user study (n = 109), we found that 19.3\% of the participants are unwilling to use any usage limiting application because a) they do not want their smartphone activities to get restricted or b) those applications are annoying. Following that, we devised a hypothesis to minimize smartphone usage among undergraduates. Finally, we designed a prototype for android, \say{App Usage Monitor} and conducted a 3-week experiment through which we found proof of concept for our hypothesis. In our prototype, we combined techniques such as nudge and visualization to increase self-awareness among the user by leveraging notification.
\end{abstract}

\begin{IEEEkeywords}
HCI, Smartphone Addiction, Smartphone Overuse, Excessive Smartphone Use, Smartphone Usage Behavior
\end{IEEEkeywords}

\section{Introduction}
With the advancement of smartphone technology, people's engagement with it is increasing proportionally. In almost every aspect of one's life, smartphones are being used, such as exchanging opinions, communicating with friends and relatives, getting news and updates about any sector. Thus, smartphone overuse is rarely generated by the phone or tablet itself, but rather the ability to instantly connect us to the games, applications, and online worlds.

Researchers have been designing various usage controlling applications to tackle this problem. These applications mainly use strategies such as blocking, nudge, visualization, and cognitive burden. But despite the applications being effective, in most cases, the user experience is not very pleasant. In a study, participants reported that restriction on smartphone use causes frustration and irritation due to potential inconveniences \cite{b1}. Okeke et al.\cite{b2} nudged the user by using repetitive vibration to reduce their smartphone use. Even if they successfully reduced digital consumption, the majority (26/31) of the participants found the vibrations to be irritating during their experiment.

\begin{figure}[htbp]
\centerline{\includegraphics[width=5cm,height=10cm]{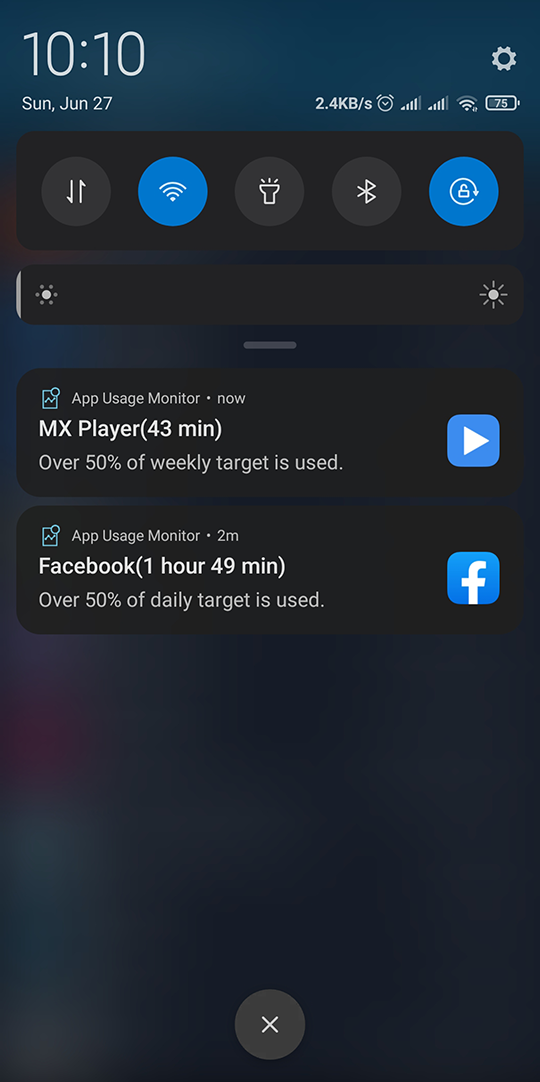}}
\caption{Usage notifications}
\label{fig1}
\end{figure}

In this paper, we hypothesized that people are not aware of their own smartphone use. Based on the hypothesis, we introduced a subtle and familiar, yet effective intervention approach. We developed an android prototype  that uses notifications to deliver useful information about all individual application's respective usage that are installed on user's device. Notifications are delivered based on the user-specified target for individual applications installed on user's smartphone (Fig.~\ref{fig1}). The notifications serve another important purpose. If a notification is selected, a detailed usage graph is shown, which helps the user increase self-awareness as self-reflection has been proved to be very successful in mitigating technology overuse. To better understand the design requirements for our prototype, we first performed an online preliminary survey with 109 undergraduate participants (68.8\% male, 31.2\% female). Then, with 16 undergraduate volunteers (81.25\% male, 18.75\% female), we conducted 3-week field research. The result from the experiment showed that notification has the potential to be a powerful weapon against excessive smartphone use. The combined result of our preliminary survey and experiment also revealed the lack of self-awareness among undergraduate students.

\section{Background and related works}
According to the techniques used by the researchers, we divided the existing works into the following categories-
\subsection{Blocking}
Such works depend on limiting user interaction with the device by restricting certain or all apps and activities. In AppDetox \cite{b3}, users can create rules for individual applications and restrict their access according to the rule. Let's Focus \cite{b4} follows a location-based approach where interaction with the smartphone is blocked within a classroom. Lock n'Lol \cite{b5} and NUGU \cite{b6} restricts a group of people from using their smartphones to increase concentration and self-regulation within the group.
\subsection{Burden}
In recent years, burdening has gained some popularity among researchers. The goal here is to discourage unnecessary smartphone using by making the user give extra effort to do certain activities on their smartphone. Interaction Restraint \cite{b7} and LocknType \cite{b8} make the user type a long number before launching certain applications on their smartphone. Time Off \cite{b9} shows animation on the top screen with notification and vibration during smartphone overuse. Kim. et al. \cite{b10} uses a lockout system that only unlocks when the reason for smartphone using is stated.
\subsection{Nudge}
A Nudge is an easy and cheap to avoid intervention that can change people’s behavior in a predictable way without restricting their choices. Good Vibrations \cite{b2} uses repetitive vibration as a nudge to alert the user during smartphone overuse. MyTime \cite{b11} is another application that displays a popup on the screen when the user-defined usage limit is crossed. Aditya et al. \cite{b12} designed a desktop browser extension that nudges the user to control social media addiction.
\subsection{Visualization}
Using visualization to make the user more self-aware has been proved to be very successful during past studies. TILT \cite{b13} is an android application that uses a graph to deliver overall smartphone usage and the number of unlocks to the user. LockDoll \cite{b14} is an Arduino-based doll that waves its hands to alert the connected group of people about their smartphone use. TimeAware \cite{b15} and meTime \cite{b16} are desktop applications designed to displays how a user spends their time across different applications to increase awareness.

In this study, we combined both nudge and visualization techniques to create an effective intervention. We designed a working prototype where notification is used to deliver real-time usage information about individual applications present on the smartphone. The prototype also has a detailed usage graph for individual applications with past information to make it easier for the user to self-reflect and minimize their smartphone activities.

\section{PRELIMINARY STUDY}
\subsection{DESIGN}
We performed a preliminary study to better understand smartphone usage behavior and mindfulness among undergraduate students. We conducted an online survey with several close-ended and semi-close-ended questions (with the prepopulated answers, there were also options to give participant's own opinions). The survey was made using Google Form. The first question of the survey is the unique university registration/id number so that we can identify them after our field experimentation phase. Besides that, there were three types of questions on the survey. At first, we asked the participants about the understanding of their own smartphone use. After that, we inquired about limiting smartphone usage, and the last part of the survey contained questions about their smartphone usage pattern and behavior, which will later be cross-checked with the actual usage data collected via \say{App Usage Monitor} to realize the users' insight about their own smartphone use. By posting the survey in several Facebook groups consisting of undergraduate students we were able to gather 109 participants who freely completed our survey (68.8\% male, 31.2\% female), age ranging from 20 to 26.
\subsection{RESULTS}
In our survey, the majority of the participants (69.7\%) felt that they overuse their smartphones. Even though 69.7\% of participants reported positively, 38.5\% said they did not even try to limit their smartphone usage when asked how they limit(or try to limit) their smartphone usage. 

Staying away from the phone is the most popular way to limit smartphone use unless we consider the responses of participants who didn't even try to limit. Among all of the participants, only 20.2\% of the participants ever used any usage-limiting app. Besides not knowing about the usage-limiting apps and just not using them, the primary issue that kept them from using any usage limiting app is that they do not want their smartphone usage to get restricted. Upon asking the participants their reason for not using any usage limiting app anymore, who previously used limiting apps, they replied that those apps were annoying and not practical. Then we ended our survey with some questions that focused on participants' insight into their own smartphone use. 62.4\% answered that they spend most of their time on smartphones doing social activities and 47.7\% answered they use Facebook the most. The second dominating app according to the response was Youtube (27.5\%). 42.2\% of the participants think that they spend more time on their smartphone at Evening (6pm-9pm).

Even though most participants did not use or know about any usage limiting app, a fair number of participants gave their negative thoughts about restrictive measures and thought them to be annoying. Even though restrictive interventions have proven to be quite successful in mitigating excessive smartphone usage, the purpose is lost if nobody uses them. In order to overcome these limitations, we have come to the following design objectives towards an optimal intervention method:
\begin{itemize}
    \item Should not interrupt the current workflow on the smartphone
    \item Should not be annoying or burdening
    \item Should not forcefully restrict any smartphone activity
    \item Should be subtle yet effective
    \item Should work with little user interaction
\end{itemize}

\section{APP USAGE MONITOR DESIGN}
Because of the ongoing popularity of dark themes and for being easy on the eyes, we have built the UI of \say{App Usage Monitor} by using dark color palettes. Every dialog inside the app also uses a custom dark theme.

\begin{figure}[hbp]
\centerline{\includegraphics[width=5cm,height=10cm]{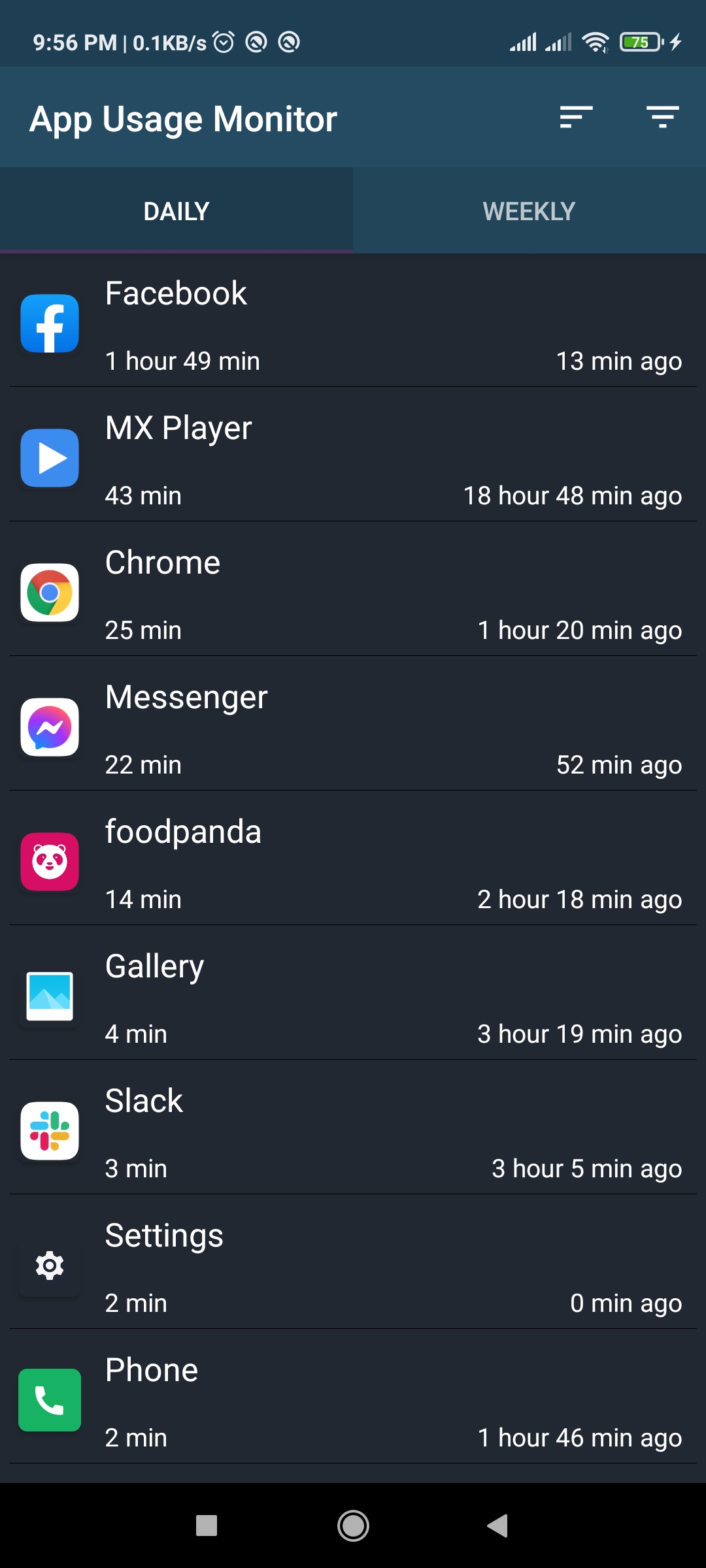}}
\caption{Main Activity}
\label{fig2}
\end{figure}

\begin{figure}[htp]
\centerline{\includegraphics[width=5cm,height=10cm]{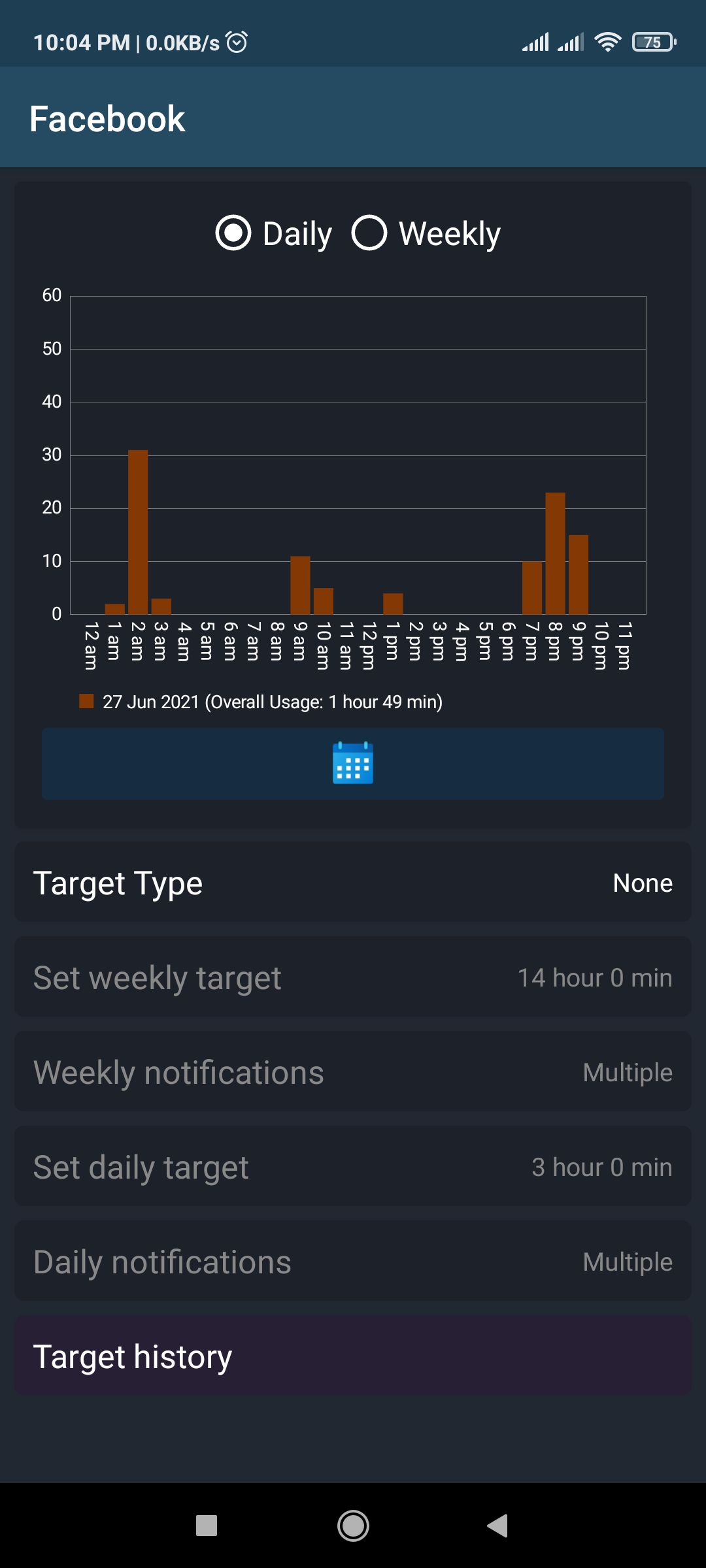}}
\caption{App Details Activity}
\label{fig3}
\end{figure}

After opening the prototype, a list of all the applications present on the smartphone along with the usage time and last access time is shown (Fig.~\ref{fig2}). Upon selecting an app from the application \say{App Details} activity is launched. In this activity (Fig.~\ref{fig3}), a detailed view of usage and all the target and notification configuration is shown to the user. This is normally fully customizable, but for the sake of the experiment, we will automatically set all the parameters to automate the experiment.

By clicking on the \say{Target History} button at the bottom of the \say{App Details} activity, the user can see a list of all the previous targets and usage for that application.


\section{FIELD EXPERIMENT}

\subsection{EXPERIMENT DESIGN}
Over the course of three weeks, we performed a controlled experiment with 16 undergraduate students to validate our hypothesis. The first week functioned as the baseline week, during which no intervention was used. We provided use notifications in the second week and tracked how smartphone usage patterns changed relative to the baseline week. The third week's goal was to examine changes in smartphone usage behavior when usage notifications were no longer sent.

When the prototype is first installed in the target device, the 0\textsuperscript{th} week starts. The reason for starting with the 0\textsuperscript{th} week is to wait for the next Sunday to come which marks as the first or baseline week. Sunday is the default start day of the week in android studio, hence we chose this day to make the experiment more consistent across all participants and to make it easier for us to monitor.
During the whole experimentation period, we disabled manual target and notification settings. All the target and notification changes automatically based on the usage and the phase of the experiment they were in. In the 0\textsuperscript{th} or baseline week, the targets were set to none, so, the participants never received any notification in the baseline week.

At the intervention week, daily and weekly targets for individual applications were automatically set. The daily target was calculated by simply taking the average of that particular application's usage and the weekly target was 7 times the daily target. There was a lower limit of 1-hour and an upper limit of 4-hour for the target so that users wouldn't get too many or too few notifications. All the notifications were also checked so that the user starts getting notifications from 50\% usage of the target.

And in the final week, all the targets and notifications were turned off to observe the participants' usual usage behavior after the intervention period.

\subsection{DEMOGRAPHIC}
Based on the criteria we have discussed in the earlier section, we were able to gather 33 (26 males, 7 females) volunteers who were willing to participate in the experiment at the beginning. However, not all of them completed the whole experiment. Some participants turned the usage access off for App Usage Monitor after the beginning of the experiment due to privacy concerns, some participants also uninstalled the app for the same reason. In some devices, the alarm repeatedly stopped triggering by the system due to power saving mode, reboot, ram management, and other android features hence the usage data could not be collected from those participants. One participant reset his device mid-way through the experiment therefore the experiment remained unfinished for him. There was also a case where a quarter of a day's worth of data was missing so we had to give up on that one too. In the end, 16 participants (13 males, 3 females) remained who completely cleared the experiment. All the participants were Bangladeshi, and their average age were ranging from 20 to 26 years.

\subsection{DATA COLLECTION}
Because of the alarm manager as well as background service on android being inconsistent, there were many irregularities on the dataset. There were some cases where the alarm did not trigger even once on the starting day of the week, so the experimentation week remained unchanged resulting in non-equal days in the experimentation weeks. Some participants baseline week, intervention week, and review week lasted longer than 1 week. So, before performing the analysis, we had to do some post-processing to make our dataset ideal for evaluation. We took the last 7 days of usage data from the baseline week and the first 7 days of usage data from the intervention and review week and filtered out the rest.

We also interviewed each of the participants who completed the trial and asked them about their opinion about the experiment and gave them a bar of big chocolate as a token of our gratitude for helping us with our research by providing valuable information.

\section{RESULT ANALYSIS AND DISCUSSION}
Over the 3 weeks of the experiment study, we collected smartphone usage data from 16 participants. By analyzing the usage log and comparing them with the survey response, we found strong evidence in favor of our hypothesis and some interesting usage patterns among undergraduate students. We divided our findings into three following parts.
\subsection{USAGE NOTIFICATION REDUCES SMARTPHONE USAGE}
\begin{figure}[htbp]
\centerline{\includegraphics[width=6cm,height=10cm]{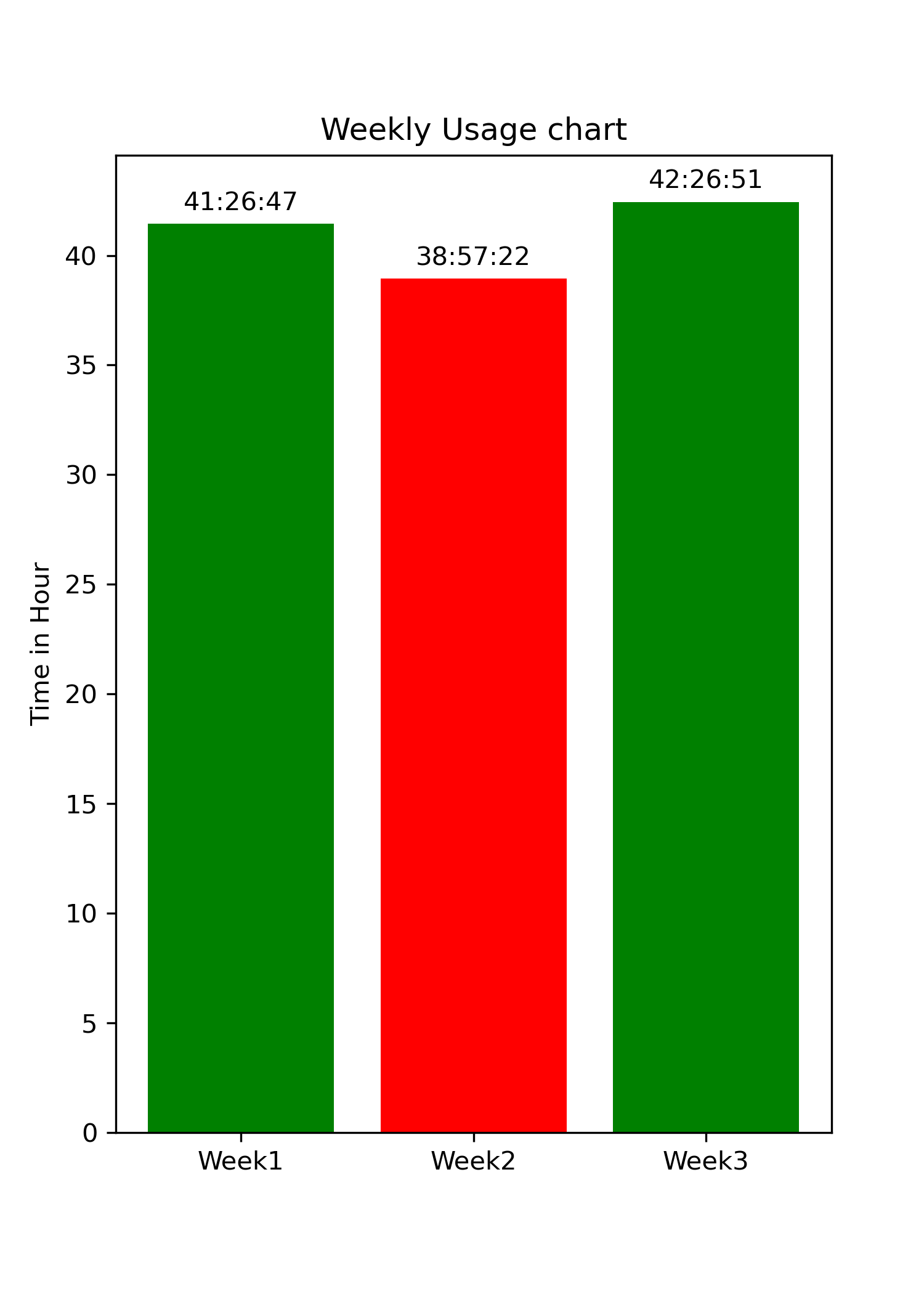}}
\caption{Usage Bar Graph of Three Experiment Week}
\label{fig4}
\end{figure}

Our results indicate that the time that participants spent on the smartphone was substantially decreased by providing app usage information via notification. 75\% of our participants' usage decreased in the intervention week compared to that of the baseline week. As we can see from Fig.~\ref{fig4}, in the baseline week, the average weekly usage per participant was 41 hours 26 minutes and 47 seconds. However, after receiving notifications on the intervention week, participants' average usage went down to 38 hours 57 minutes and 22 seconds leading to an overall usage reduction of 6.01\%. Nevertheless, when the usage notification stopped showing in the review week, the average weekly usage went up to 42 hours 26 minutes and 51 seconds (8.23\% increase compared to intervention week) which is almost as equal to the average usage time of the baseline week that further assists our hypothesis. Fig.~\ref{fig4} shows the overall smartphone use across three experiment weeks.

\subsection{LOW SELF-AWARENESS}
By comparing the survey response with the actual usage log collected from the participants who completed the 3-week study, we found a clear lack of self-awareness among the undergraduate smartphone users. For the comparison to be fair, we only sampled the usage data from the baseline week since there was no intervention, and the main purpose of the baseline week was to be the point of reference for the other experimentation weeks. After analyzing the survey response and the usage log, we found, among all the participants, 62.5\% had higher average daily usage than their estimated average daily usage on the survey. Half (50\%) of the participants were totally wrong about the application they used the most on their smartphones. 56.25\% participants’ estimation about the parts of the day (Table~\ref{tab1}) they use their smartphone the most from the survey did not match their actual usage history thus implying that not only the undergraduate student lacks self-awareness about their own smartphone usage but also not having a good understanding about their overall smartphone usage behavior.

\begin{table}[hb]
\caption{Parts of the Day}
\begin{center}
\begin{tabular}{|c|c|}
\hline
 Neme & Time Range \\
\hline
Morning & 6am-12pm \\
Afternoon & 12pm-6pm \\
Evening &  6pm-12am \\
Late at Night & 12am-6am \\
\hline
\end{tabular}
\label{tab1}
\end{center}
\end{table}

At the post-experiment interview, when shown the app usage statistics of their past three weeks, most of the participants were surprised about their usage being higher than they would imagine. One participant quoted, \say{I couldn’t even imagine that I use my smartphone for over 6 hours a day on average}. And when we showed them the number of times they opened their most launched app, some of the participants thought that the data we presented them was fake at first. These reactions from the participants further reveal that the self-awareness among the undergraduate students about using their smartphones is indeed low.
\subsection{SMARTPHONE USAGE BEHAVIOR}
\begin{figure}[htbp]
\centerline{\includegraphics[width=0.5\textwidth]{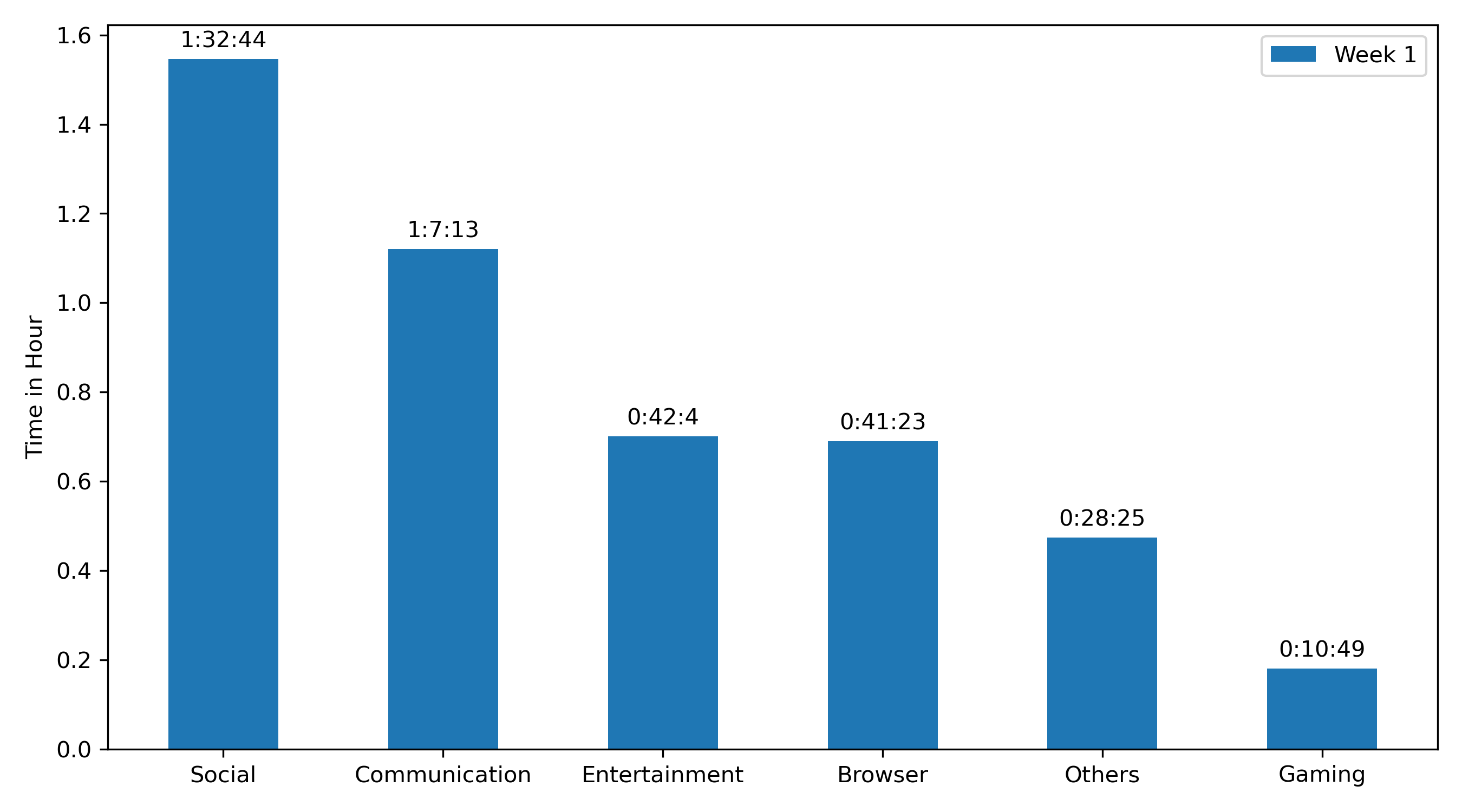}}
\caption{Categories based on Users' Average Daily Application Use}
\label{fig5}
\end{figure}

\begin{figure}[htbp]
\centerline{\includegraphics[width=0.5\textwidth]{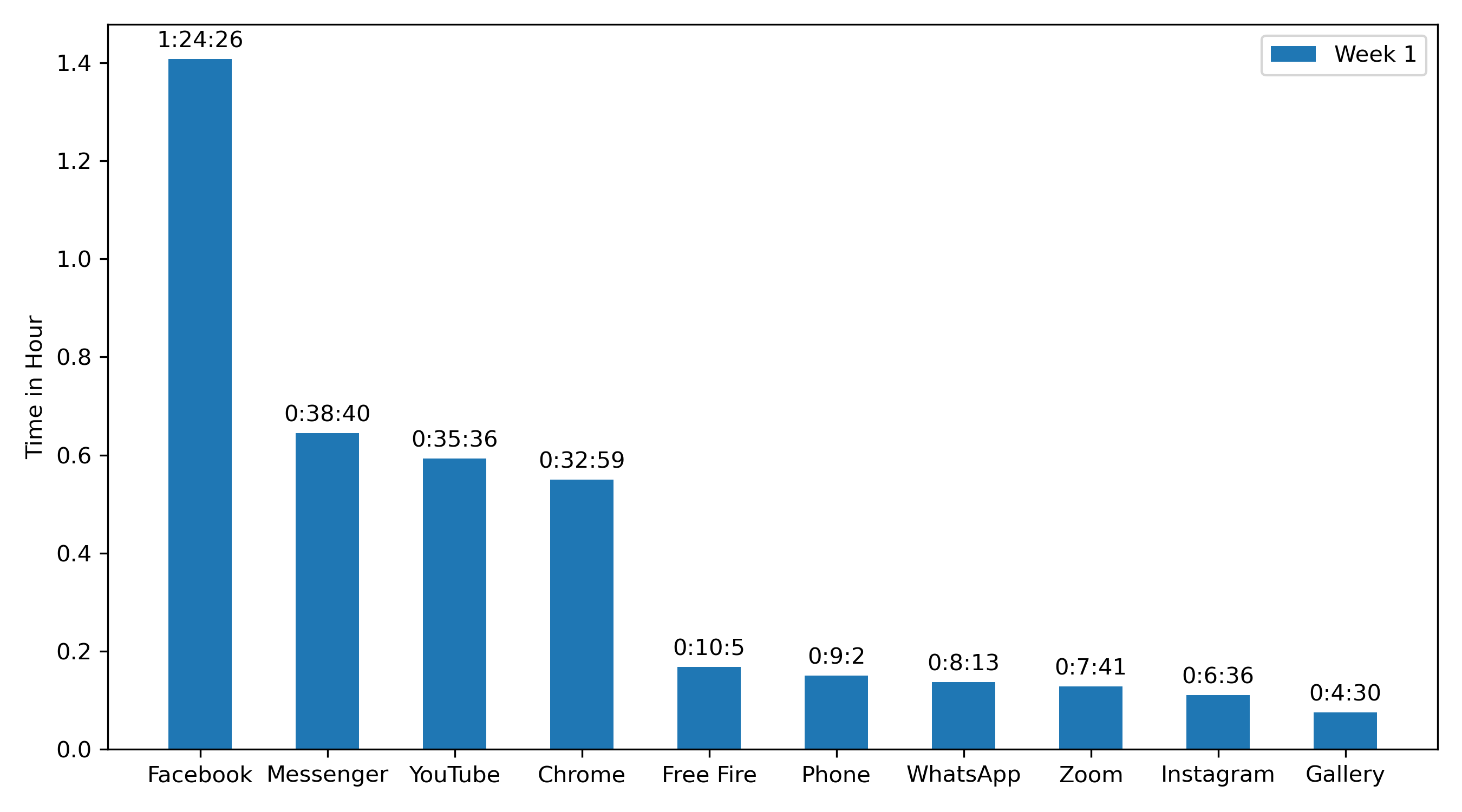}}
\caption{Top 10 Used Applications in First Week (Average Daily Use Time Based)}
\label{fig6}
\end{figure}

\begin{figure}[htbp]
\centerline{\includegraphics[width=0.5\textwidth]{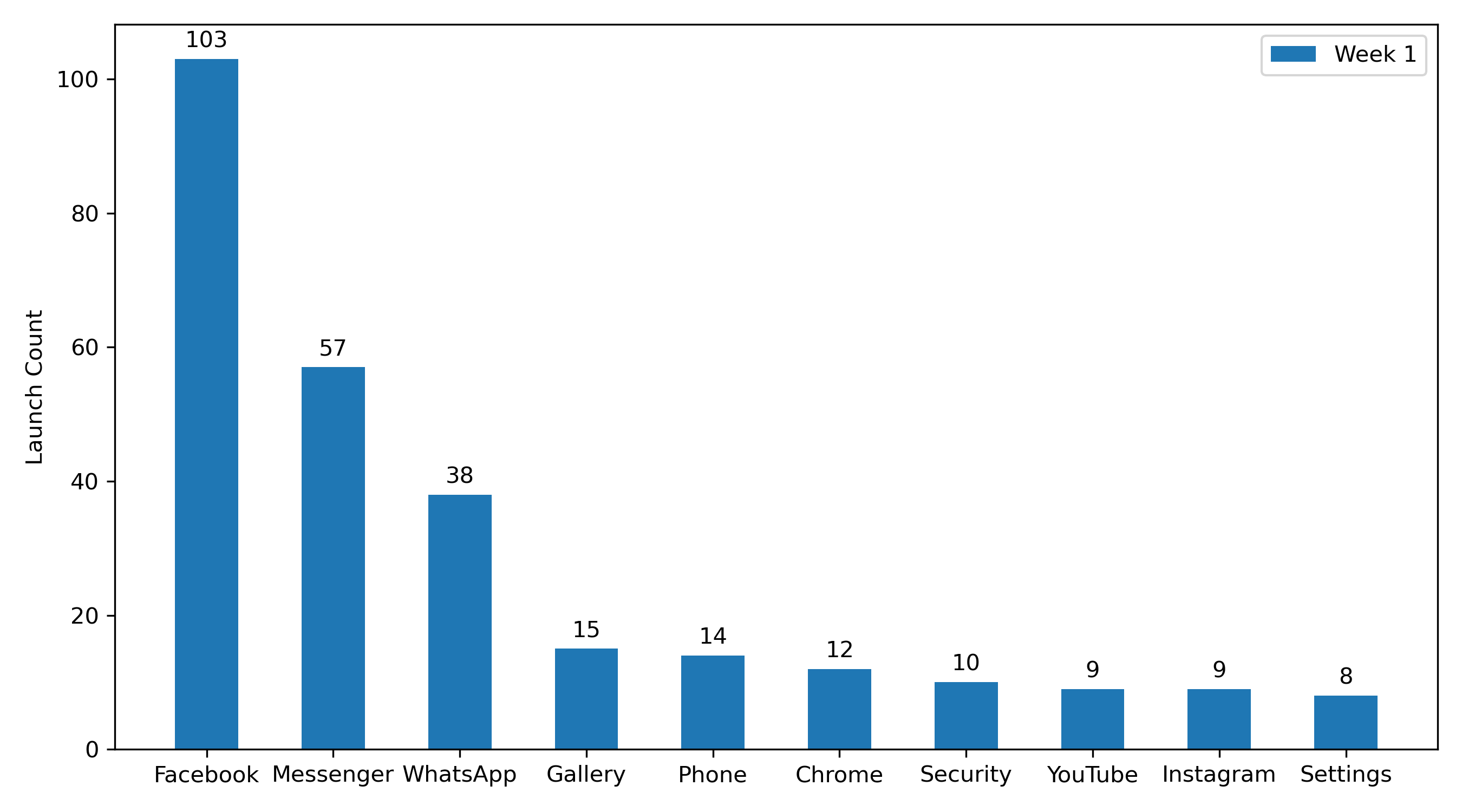}}
\caption{Top 10 Used Applications in First Week (Average Daily Launch Count Based)}
\label{fig7}
\end{figure}

We found a wide range of application usage among our participants in our study. However, among all of these apps, social media apps were the most popular among undergraduates. Fig.~\ref{fig5} shows all of the most popular categories together with their average daily usage time. Fig.~\ref{fig6} depicts the participants' most frequently used programs based on their average daily usage. Facebook was the most popular app, with about 1 hour 24 minutes and 26 seconds of average daily usage time. The most launched application among our participants was also Facebook (Fig.~\ref{fig7}).

\begin{figure}[htbp]
\centerline{\includegraphics[width=0.5\textwidth]{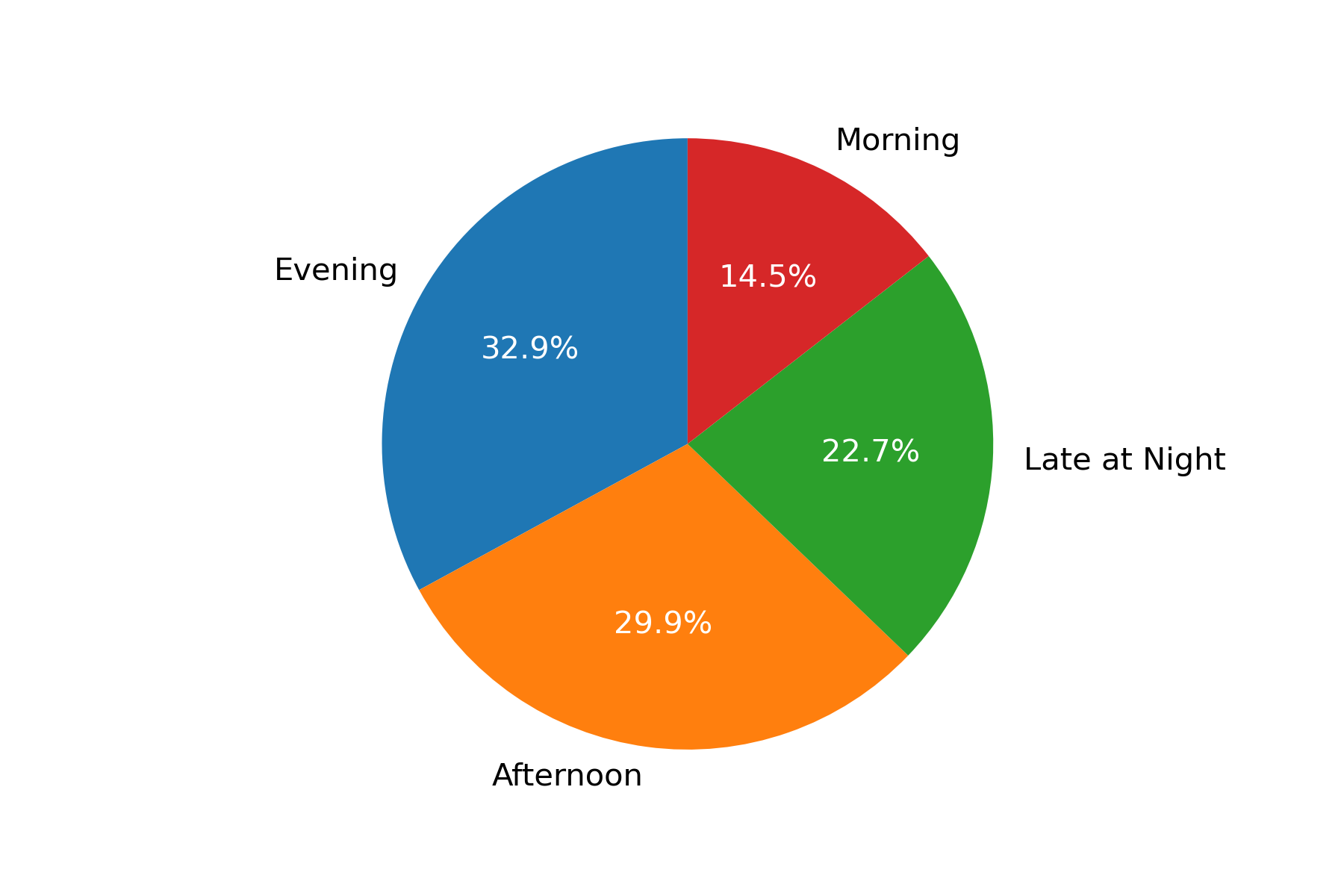}}
\caption{Smartphone Using Frequency According to Parts of the Day}
\label{fig8}
\end{figure}

In our study, we discovered that undergraduate students use their smartphones the most at \say{Evening} and the least at \say{Morning}. Fig.~\ref{fig8} illustrates smartphone using frequency according to parts of the day (Table~\ref{tab1}).

\section{DESIGN RECOMMENDATIONS}
The present intervention design's major flaw is that it is very aggressive. Many smartphone use mitigation technologies restrict access to applications and specific activities in order to minimize smartphone usage. To discourage the user from using smartphones, several tools utilize cognitive load to add extra effort before using the smartphone. This type of intervention might work in a controlled experiment environment where participants are required to use the mitigation tool for the whole experiment duration and give off promising results. However, if individuals do not use the end product in their everyday lives, the goal is defeated.

Some methods aren't forceful, but they're so quiet that the user doesn't even notice they're there. Users must directly interact with these applications in these circumstances in order to reap the benefits. Google's Digital Wellbeing project is an excellent example of this type of effort.

So, before designing smartphone usage mitigation systems, the researcher must find the sweet spot between restrictiveness and silence and go on from there. In this study, we showed that subtle and familiar interventions like notification can be leveraged to create a viable smartphone mitigation tool. We believe that additional studies should employ similar intervention approaches to avoid user dissatisfaction and annoyance. In this age of ease, rather than attempting to reduce smartphone overuse by making smartphone usage unpleasant, we should investigate how we might make smartphone use more comfortable while still controlling excessive smartphone use.

\section{CONCLUSION}
In this article, we conducted comprehensive research on how we can alleviate the worldwide issue of excessive smartphone usage among undergraduate students. Initially, we investigated the consequences of excessive smartphone use. We saw how excessive smartphone use impacts our job, social, and personal lives, as well as our physical and mental health. We reviewed several previous works and conducted a user analysis in which we surveyed 109 undergraduate students to better understand their smartphone use habits. Based on the literature review and user study, we hypothesized the lack of self-awareness among the undergraduate students and proposed a notification-based smartphone mitigation prototype \say{App Usage Monitor} to minimize excessive smartphone usage among undergraduate students. To prove our hypothesis and understand smartphone usage behavior, we conducted a three-week-long experiment with 16 volunteers.

The result revealed that most of the participants' opinion about their own smartphone usage was wrong which verifies our hypothesis.The result of the experiment also showed a significant decrease in smartphone usage among the participants when the intervention was applied compared to their usual usage. So we can say that making user more self-aware reduces excessive smartphone use.

Finally, we think subtle and familiar interventions can be used to create powerful tools in mitigating various technology overuse. We hope to see more works that try to achieve smartphone usage mitigation via unforced techniques.

\bibliography{IEEEabrv,ref_file} 
\bibliographystyle{IEEEtran} 

\end{document}